\documentclass[reprint,aps,pra,twocolumn,showpacs]{revtex4-2}
\usepackage{amssymb}
\usepackage{graphicx}
\usepackage{dcolumn}
\usepackage{bm}
\usepackage{amsmath}
\usepackage{soul,color}
\usepackage{textcomp}
\usepackage{times}
\usepackage{csquotes}
\usepackage{booktabs}
\usepackage{xcolor}
\usepackage{mathrsfs}

\usepackage[colorlinks,linkcolor=red,citecolor=cBlue]{hyperref}
\providecommand{\U}[1]{\protect\rule{.1in}{.1in}}
\definecolor{cGreen}{RGB}{0,0,0}
\definecolor{cBlue}{RGB}{45,51,180}
\definecolor{cmagenta}{RGB}{205,0,100}

\usepackage{ulem}
\usepackage{xcolor}

\begin{document}
	
	\title{Partial confinement in a quantum-link simulator}
	
	\author{Zheng Tang$^1$}
	
	\author{Fei Zhu$^1$}
	
	\author{Yi-Fan Luo$^{2,3}$}
	
	\author{Wei Zheng$^{2,3,4}$}
	\email{zw8796@ustc.edu.cn}

	\author{Li Chen$^1$}
	\email{lchen@sxu.edu.cn}
	
	\affiliation{
		$^1${Institute of Theoretical Physics, State Key Laboratory of Quantum Optics and Quantum Optics Devices, Shanxi University, Taiyuan 030006, China}\\
		$^2${Hefei National Laboratory for Physical Sciences at the Microscale and
			Department of Modern Physics, University of Science and Technology of China,
			Hefei 230026, China}\\
		$^3${CAS Center for Excellence in Quantum Information and Quantum Physics,
			University of Science and Technology of China, Hefei 230026, China}\\
		$^4${Hefei National Laboratory, University of Science and Technology of China,
			Hefei 230088, China}
	}
	
	\begin{abstract}
		Confinement/deconfinement, captivating attributes of high-energy elementary particles, have recently garnered wide attention in quantum simulations based on cold atoms. Yet, the \textit{partial confinement}, an intermediate state between the confinement and deconfinement, remains underexplored. The partial confinement encapsulates the phenomenon that the confining behavior of charged particles is contingent upon their relative positions.
		In this paper, we demonstrate that the spin-1 quantum link model provides an excellent platform for exploring partial confinement. We conduct a comprehensive investigation of the physics emerging from partial confinement in both the context of equilibrium and non-equilibrium dynamics. Potential experimental setups using cold atoms are also discussed. Our work offers a simple and feasible routine for the study of confinement-related physics in the state-of-the-art artificial quantum systems subject to gauge symmetries. 
	\end{abstract}
	
	\maketitle

	\section{Introduction}
	Confinement is a fundamental property prominently observed in quantum chromodynamics (QCD), where the inter-quark potential increases with their distance \cite{PhysRevD.10.2445,allan1977mechanism,greensite2011introduction}. This prevents the existence of isolated quarks due to energetic instability; instead, they prefer to bind together into hadrons, either as mesons (quark-antiquark pairs) or baryons (triplets of quarks). Although the concept originated in QCD, analogous phenomena can also manifest in strongly coupled charges in quantum electrodynamics (QED) \cite{schwinger1962gauge,coleman1976more}, i.e., the charge confinement. 
	Dimensional analysis indicates that the dimensionality of the coupling constant is determined by the dimensions of the system. Specifically for (3+1)D, the coupling constant is dimensionless, leading to the deconfined Coulomb potential $\sim 1/r$, where $r$ is the distance between two charges. The deconfinement-confinement phase transition can occur by tuning the coupling strength and the temperature \cite{svetitsky1986symmetry}. However, for (1+1)D QED, also known as the Schwinger model, the dimensionality of the coupling constant scales linearly with $r$. Consequently, apart from certain exceptional cases, the confining phase becomes quite prevalent.
	 Furthermore, confinement and deconfinement phenomena also appear in emergent gauge theories from strongly-correlated electrons and recent developed Rydberg atomic arrays \cite{cheng2024emergent,Cheng_2023,PhysRevResearch.4.L032037,surace2020lattice,PRXQuantum.4.020301}. For instance, transition between valence bond solid to spin liquid phase can be understood in a picture of confinement-deconfinement transition of spinons \cite{RevModPhys.75.913,PhysRevLett.114.017203}. However, large scale numerical investigation of the real time dynamics of confinement or deconfinement on classical computers is challenging.
	 
Recently, much effort has been made to overcome this barrier through quantum simulation \cite{wiese_ultracold_2013, banuls_simulating_2020, zohar_quantum_2016,XXQS202301007, PRXQuantum.4.027001,doi:10.1098/rsta.2021.0069,PhysRevLett.125.030503,halimeh2020fatelatticegaugetheories,PhysRevResearch.5.043128,doi:10.1098/rsta.2021.0064,Klco_2022} which leverages systems with discrete degrees of freedom. This includes analog simulations using optical lattices \cite{PRXQuantum.4.020330,Yang2020NatureObservation,mil_scalable_2020,zhou2022thermalization,wang_interrelated_2023,zhang2023observation,González-Cuadra_2017,Homeier2023,Tagliacozzo2013,halimeh2023coldatomquantumsimulatorsgauge,Zache_2018,PhysRevLett.112.120406,zheng2021floquetengineeringdynamicalz2,PhysRevLett.110.125304,PhysRevA.105.023322,doi:10.1126/sciadv.aav7444,Kuno_2015,PhysRevResearch.2.033361,Goldman_2014,PhysRevLett.109.175302,PhysRevLett.107.275301,PhysRevLett.110.055302} or trapped ions \cite{PhysRevResearch.2.023015,hauke2013quantum,Bazavan2024,Blatt2012}, and digital simulations realized on various quantum computing platforms \cite{PhysRevA.95.023604,PhysRevA.98.032331, PhysRevD.101.074512, PhysRevA.73.022328, PhysRevD.102.094501, Muschik_2017,kokail_self-verifying_2019,martinez_real-time_2016,Davoudi_2021,PhysRevD.106.054508,kan2022latticequantumchromodynamicselectrodynamics,Shaw2020quantumalgorithms,PhysRevA.62.022311}.

The quantum link model (QLM) \cite{CHANDRASEKHARAN1997455} serves as one of the most commonly used approaches to simulate lattice gauge theories, which are based on the Hamiltonian formalism with space discretized while time remains continuous. In QLMs, matter particles are placed on lattice sites, while gauge spins with a finite local Hilbert space are located on the links connecting neighboring sites.
The realization of QLM is considered a powerful approach for exploring strongly coupled QED, as strong coupling renders perturbative field theory ineffective, and hence quantum simulation can essentially circumvent the issues encountered by classical simulations, such as the sign problem in quantum Monte Carlo \cite{troyer2005computational}. In these quantum-link simulators, both confinement and deconfinement have been extensively studied, encompassing theoretical \cite{PhysRevLett.109.175302,cheng2022tunable,BingYangspin-1/2tuningtheta,zohar_simulating_2012,cheng2023gauge,PhysRevX.6.041040,qi_gauge_2024,PhysRevB.109.245110,PhysRevLett.127.167203,PhysRevLett.107.275301} and experimental \cite{mildenberger2022probing,zhang2023observation} contexts. Particularly for the spin-1/2 QLM, the confinement-deconfinement transition has been experimentally signified in dynamics through tuning the topological angle \cite{zhang2023observation}.
	
	In this paper, we delve into an intermediate phenomenon between confinement and deconfinement in (1+1)D QED, called \textit{partial confinement}, which has hitherto remained unexplored within the context of quantum simulation. It refers to the situation where the confining or deconfining status between charges depends on their relative positions. Our study draws inspiration from the seminal work of S. Coleman in the 1976 \cite{coleman1976more},  which studied half-asymptotic particles in the continuum Schwinger model.
	Here, we demonstrate that the 1D spin-1 QLM can serve as an excellent platform for observing partial confinement: it retains the essential physics while being simple enough to be realized within the scope of current experimental capabilities. Taking the spin-1 QLM as a background, we introduce the basic concept of partial confinement and discuss the associated emergent physics, both in equilibrium and non-equilibrium dynamics.
	
	It is worth clarifying that the terminology of partial (de)confinement has already been introduced in QCD \cite{Hanada2019,Hanada2023, Gautam2023}. 
	Therein, partial confinement refers to a phase where color degrees of freedom split into confined and deconfined sectors, with a subgroup of the SU(N) gauge group becoming deconfined while the remainder stays confined. This is typically characterized by a non-uniform distribution of Polyakov loop phases and the N-dependence of thermodynamic quantities.
	As such, this notion of partial confinement carries a different physical meaning compared to ours defined above.
	
	The rest of this paper is structured as follows: Sec.~\ref{QLM} provides a review of the spin-1 quantum link model, detailing its essential physical features. In Sec.~\ref{PCEquilibrium}, we delve into the equilibrium properties of partial confinement within the spin-1 QLM. Sec.~\ref{PCDynamics} discusses how partial confinement manifests in non-equilibrium dynamics. In Sec.~\ref{ExperimentalDiss}, we explore the feasibility of experimental realizations using cold atoms trapped in optical super-lattices. A brief summary can be found in Sec.~\ref{Conclusion}.
	
	\section{Spin-1 Quantum Link Model}	\label{QLM}
	
\begin{figure}[pbt]
		\begin{center}
			\includegraphics[width=.47 \textwidth]{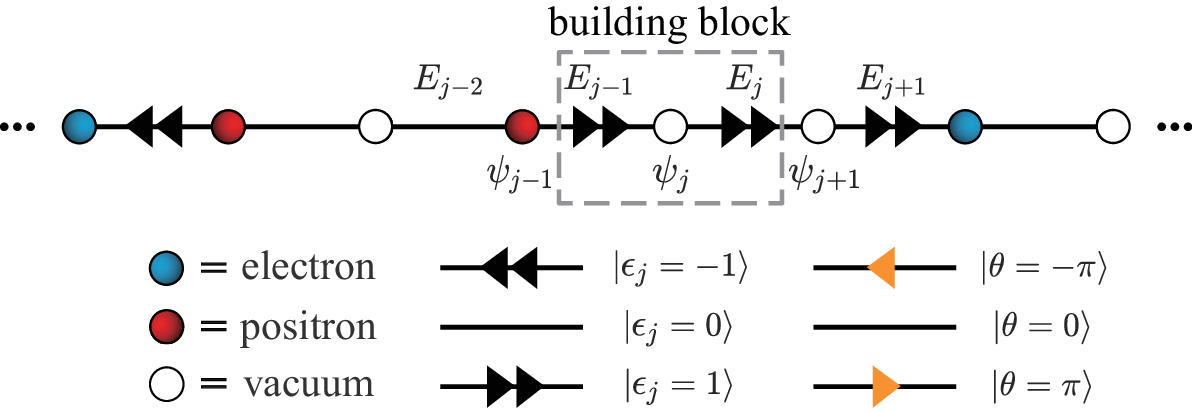}
		\end{center}
		\caption{Upper panel: The schematic representation of a spin-1 quantum link model in the particle-antiparticle picture. Circles denote matter fields residing on lattice sites; links between neighboring sites represent electric fields $E_j$ realized by gauge spins $S_j^z$. Two adjacent gauge spins with a matter field in between constitute a building block. Lower panel: The notation convention for matter-field occupations and electric spin states.}
		\label{Fig1}
	\end{figure}

	The spin-1 quantum link chain is characterized by the Hamiltonian \cite{PhysRevLett.109.175302, CHANDRASEKHARAN1997455,kuhn2014quantum}
	\begin{equation}
		\begin{aligned}
			{H} =& -J \sum_{j = 1}^{N-1}\left(\frac{1}{\sqrt{2}} {\psi}_j S_j^{+} \psi_{j+1}+\text { h.c. }\right) \\
			& +m \sum_{j=1}^N {\psi}_j^{\dagger} {\psi}_j + g \sum_{j=1}^{N-1}\left[(-1)^{j+1}S_j^z+\frac{\theta}{2 \pi}\right]^2,
		\end{aligned}
		\label{H}
	\end{equation}
	where ${\psi}_{j}$ denotes the local matter fields of fermions, and ${S}_{j}^{z,\pm}$ are the spin-1 Pauli operators representing the gauge spins living on the link between two neighboring sites $j$ and $j+1$. The number of sites $N$ must be even, allowing for the division of fermions into electrons and positrons, yielding the \textit{particle-antiparticle picture}: 
	 for $j\in \text{odd}$, the unoccupied and occupied status of electrons are respectively denoted by empty circles and blue disks in Fig.~\ref{Fig1}; whereas for $j\in \text{even}$, the corresponding occupation status of positrons are illustrated by empty circles and red disks.

	The last two terms in $H$ respectively represent the fermion mass with $m\ge0$ and the electric-field energy $g\sum_j E_j^2$ with $g\ge0$ and 
	\begin{equation}
		E_j = (-1)^{j+1}S_j^z + \frac{\theta}{2\pi}.
	\end{equation}
	A local $E_j$ consists of two parts: $S_j^z$ represents the quantized electric field capable of adopting three states $|s_j = -1\rangle$, $|0\rangle$, and $|1\rangle$. The factor $(-1)^j$ indicates that the electric field $E_j$ depends on the gauge spins in an alternating manner: $E_j$ is aligning with $S_j^z$ for $j\in\text{odd}$, while they differ by a minus sign for $j\in\text{even}$. The c-number $\theta$ is called the topological angle \cite{coleman1976more,PhysRevLett.37.172,callan1976structure,t1976computation}, which reflects the influence of an external static electric field. Thereby, the eigenvalues of $E_j$ can also take three real values, i.e., $\epsilon_j = (-1)^{j+1}s_j^z + \theta/2\pi$. Restricting $E_j$ within finite status is a key advantage of the QLM, as it facilitates experimental simulation of electric fields using a finite number of discrete degrees of freedom (such as cold atoms with internal spins). In Fig.~\ref{Fig1}, the black arrows on links represent the electric-field state $\epsilon_j$ in the case of $\theta=0$, where each left(right)-pointing arrow denotes $\epsilon = -1/2(1/2)$. Orange arrows depict the background field of $\pm1/2$, corresponding to the cases of $\theta=\pm\pi$, respectively. A pair of opposing arrows on the same link can mutually cancel each other out.
	
	The first term in Eq.~(\ref{H}) characterizes the matter-gauge interaction. This term provides the Schwinger mechanism, i.e., a pair of electron and positron merge together accompanied by the emission of gauge photons, as well as its reverse process. Photon creation/annihilation is reflected in the change of spin states via $S_j^{\pm}$.
	
	The spin-1 QLM exhibits a U(1) local gauge symmetry generated by the local Gauss operator
	\begin{equation}
		\begin{aligned}
			G_j &= E_j - E_{j-1} - (-1)^{j}{\psi}^\dagger_{j} {\psi}_{j} \\
			&=(-1)^{j+1}\left[S_j^z+S_{j-1}^z + {\psi}^\dagger_{j} {\psi}_{j}\right]
			\label{G}
		\end{aligned},
	\end{equation}
	satisfying $[G_j,H] = [G_j,G_{k\neq j}]=0$. This ensures the invariance of the Hamiltonian under arbitrary U(1) gauge transformations $U_j = \exp(i \phi_j G_j)$. As per Eq.~(\ref{G}), $G_j$ is defined within a \textit{building block} consisting of two gauge fields $\{E_{j-1}, E_j\}$ and a matter field $\psi_j$ in the middle [see the box with dashed lines in Fig.~\ref{Fig1}]. The quantum number of $G_j$, denoted by $q_j$, is called the {static gauge charge},  which is apparently a good quantum number. $q_j$ characterizes the difference between the net electric flux $E_j - E_{j-1}$ and the {matter charge} $\psi_j^\dag \psi_j$. The additional factor $(-1)^j$ arises from the opposite matter charges carried by electrons and positrons. The U(1) gauge symmetry divides the total Hilbert space into several gauge sectors, each labeled by a unique set of gauge charges $\mathbf{q} = \{ q_1,q_2, ..., q_{N-1}\}$. Notably, the sector with $\mathbf{q} = \mathbf{0}$ is called the physical sector, as now Eq.~(\ref{G}) aligns with the traditional Gauss's Law in the classical electrodynamics.
	
	In some literature \cite{PhysRevLett.109.175302,huang2019dynamical,hauke2013quantum,marcos_superconducting_2013}, the QLM [Eq.~(\ref{H})] is presented in an alternative form within the \textit{particle picture}, described by the Hamiltonian with staggered mass 
	\cite{banks_strong-coupling_1976, PhysRevD.11.395}
	\begin{equation}
		\begin{aligned}
			\tilde {H} =& -J \sum_{j = 1}^{N-1}\left(\frac{1}{\sqrt{2}} {\psi}_j^\dagger S_j^{+} \psi_{j+1}+\text { h.c. }\right) \\
			& +m \sum_{j=1}^N (-1)^j {\psi}_j^{\dagger} {\psi}_j + g \sum_j^{N-1}\left(S_j^z+\frac{\theta}{2 \pi}\right)^2,
		\end{aligned}
		\label{Hf}
	\end{equation} 
	which relates to the particle-antiparticle Hamiltonian $H$ [Eq.~(\ref{H})] through a particle-hole transformation on odd sites, i.e.  $\psi_{j\in \text{odd}} \mapsto \psi^\dagger_{j\in \text{odd}}$, as well as a transformation on even gauge spin $S^{+}_{j\in \text{even}}\mapsto -S^{-}_{j\in \text{even}}$ and $S^z_{j\in \text{even}}\mapsto -S^z_{j\in \text{even}}$. In this framework, $\psi^\dagger_j \psi_j$ at odd sites represents the occupation below the Dirac sea, thereby exhibiting a negative energy (mass) $-m$. The creation of a hole $\psi_{j}$ in the particle picture is equivalent to the creation of an electron $\psi_{j}^\dagger$ in the particle-antiparticle picture. Note that both Hamiltonians, $H$ and $\tilde H$, are mathematically equivalent for calculation purposes. Therefore, we proceed with our following analysis using the particle-antiparticle picture.
	
	\section{Partial Confinement in Equilibrium}  \label{PCEquilibrium} 
	
\begin{figure}[pbt]
		\begin{center}
			\includegraphics[width=.47 \textwidth]{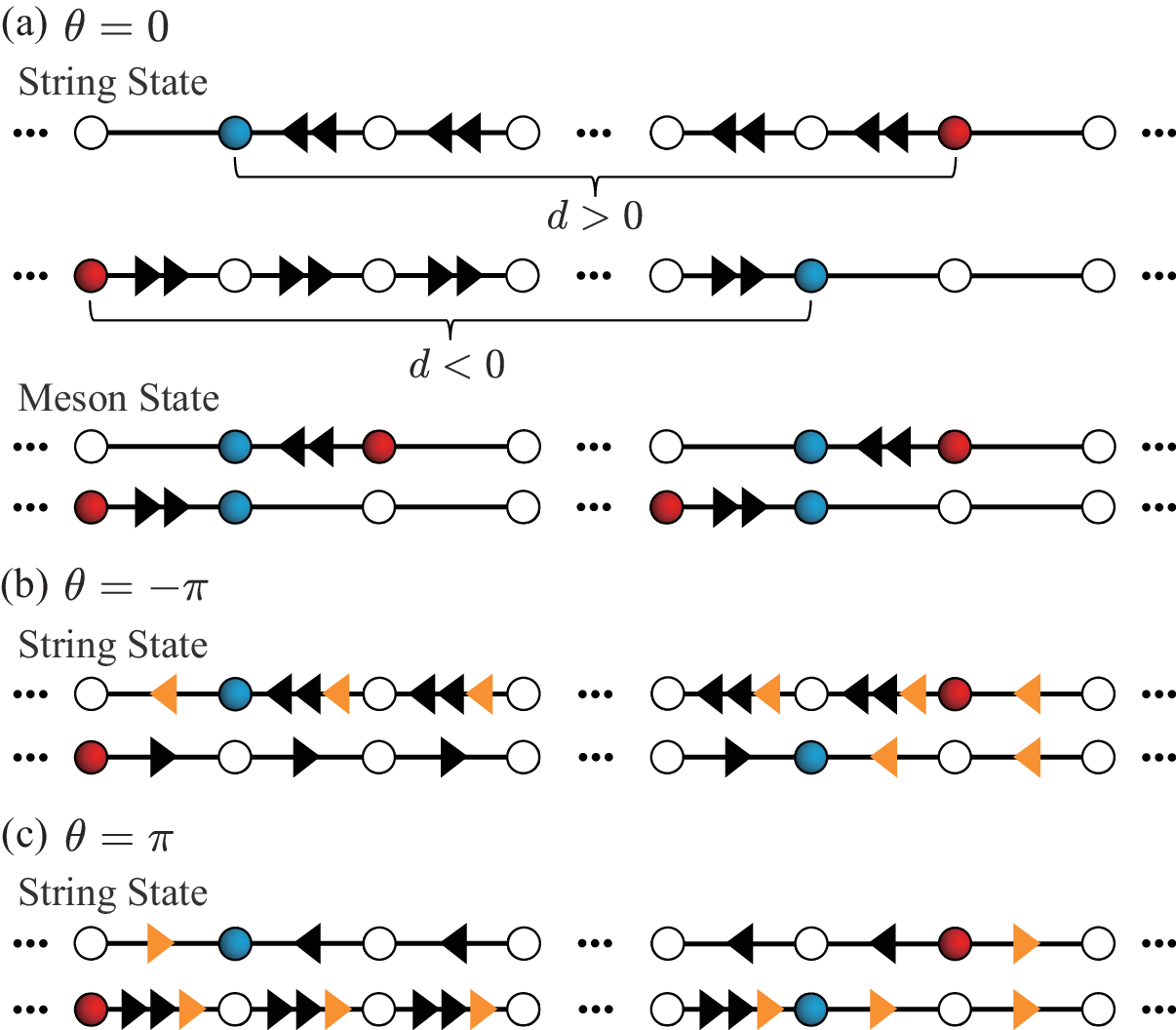}
		\end{center}
		\caption{The configuration of states at $J=0$. (a) For the case of $\theta=0$, the first and second rows depict the string states with $d>0$ and $d<0$, respectively. The third and fourth rows illustrate the corresponding configurations of the meson states. (b) and (c) correspond to the cases of $\theta=\mp\pi$, where only the configurations of the string states are shown.}
		\label{Fig2}
	\end{figure}

	The confining effects can be clearly demonstrated by the properties of equilibrium states in the physical sector $\mathbf{q}=\mathbf{0}$. We first focus on the simplest case of $J=0$, where the matter and gauge fields are decoupled, thereby $\psi^\dagger_j \psi_j$ and $S_j^z$ being conserved. We insert a pair of test electron and positron into the vacuum, separated by a distance $d$, as schematically shown in Fig.~\ref{Fig2}(a). $d$ can be positive or negative, with $d>0$ indicating the electron is to the left of the positron, and vice versa. The case of $d=0$ is excluded by the Pauli exclusion principle. According to Gauss's Law [Eq.~(\ref{G})], the system is in the string state 
	\begin{equation}
		|\psi_\text{str}\rangle = 
		\left\{ 
		\begin{aligned}
			|\cdots 0 _0 1 _{-1} 0 _{1} 0\cdots 0 _{1} 0_{-1} 1_ 0 0 \cdots\rangle ~\text{   for   } d>0 \\
			|\cdots 1 _{-1} 0 _{1} 0 _{-1} 0 \cdots 0_{-1}1_{0}0_ 0 0 \cdots\rangle \text{   for   } d<0
		\end{aligned}
		\right. ,
		\label{PsiStr}
	\end{equation} 
	where an electric string exists between the matter charges [see Fig.~\ref{Fig2}(a)]. The notation convention in a building block is $| _{s_{j-1}}, n_j, ~_{s_{j}} \rangle$, with $s_{j}$ and $n_j$ being the quantum numbers of $S^z_{j}$ and ${\psi}^\dagger_{j} {\psi}_{j}$, respectively. For a sufficiently large $m$, string state is the ground state, as $m>0$ does not favor matter-particle excitations.
	
\begin{figure*}[t]
		\centering
		\includegraphics[width= 1.0 \textwidth]{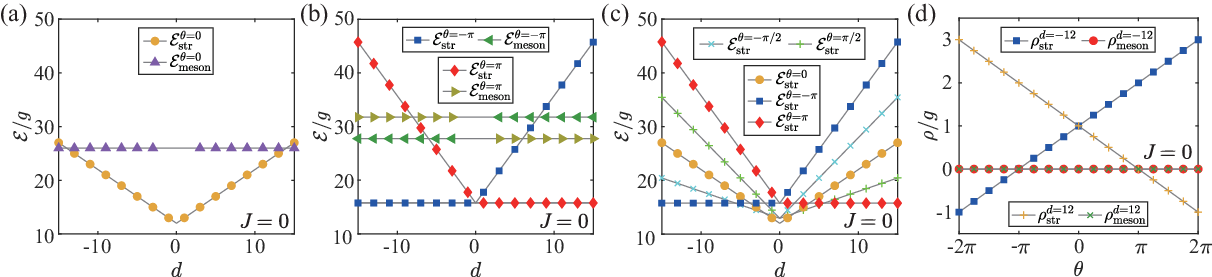}
		\caption{Equilibrium states' energy $\mathcal{E}$ and string tension $\rho$ in the case of $J=0$, with $N=16$, $g=1$ and $m=6g$ being fixed. Panel (a) displays the dependence of the string state energy $\mathcal{E}_\text{str}$ and the meson state energy $\mathcal{E}_\text{meson}$ on $d$ for $\theta=0$; (b) depicts the behavior of $\mathcal{E}_\text{str}$ and $\mathcal{E}_\text{meson}$ for $\theta=\pm\pi$; (c) shows $\mathcal{E}_\text{str}$ for different values of $\theta$; (d) presents the string tension $\rho$ as a function of the topological angle $\theta$ for fixed $d=\pm12$.}
		\label{Fig3}
	\end{figure*}

	The confining property is determined by the variance of the state energy $\mathcal{E} = \langle \psi |H| \psi \rangle$ on $d$, where the topological angle $\theta$ plays a crucial role. To be more specific, the string state has energy 
	\begin{equation}
		\mathcal{E}_\text{str}=2m+g|d|\left[-\mathrm{sgn}(d)+\frac{\theta}{2\pi}\right]^2+g\left(N-|d|-1\right)\left[\frac{\theta}{2\pi}\right]^2,
		\label{Estr}
	\end{equation}
	where $\mathrm{sgn}(x)$ is the sign function being defined as $\mathrm{sgn}(x>0) = 1$ and $\mathrm{sgn}(x<0) = -1$.
	When $\theta=0$, $\mathcal{E}_\text{str} = 2m + g|d|$ which is linearly proportional to $|d|$, irrelevant to the sign of $d$, which is a typical nature of confinement. The dependence of $\mathcal{E}_\text{str}$ on $d$ is numerically confirmed by the line with circles in Fig.~\ref{Fig3}(a). Pictorially, as depicted in Fig.~\ref{Fig2}(a), confinement manifests as an increase in the length of the string with local $|\epsilon_j|=1$, leading to an increment in the total electric energy. The string tension is defined as $\rho = \partial \mathcal{E}/\partial {|d|}$, which evaluates to $\rho_\text{str} = g$ for the string state.
	The energy instability of the string state manifests in a possible decay into the meson state with lower energy. The meson configuration is 
	\begin{equation}
		|\psi_\text{meson}\rangle = 
		\left\{ 
		\begin{aligned}
			|\cdots 0 _0 1 _{-1} 1 _0 0 \cdots 0_0 1_{-1} 1_0 0 \cdots\rangle \text{   for   } d>0 \\
			|\cdots 1 _{-1} 1 _0 0 _0 0 \cdots 1_{-1} 1_{0} 0_0 0 \cdots\rangle \text{   for   } d<0
		\end{aligned}
		\right. ,
		\label{PsiMeson}
	\end{equation} 
	as illustrated in Fig.~\ref{Fig2}(a), which results from the binding of test charges to their nearest anti-particles, with the inter-particle electric string being screened. Hence, the decay process is also called the string breaking. The meson state has an energy
	\begin{equation}
		\mathcal{E}_\text{meson}=4m+2g\left[-\mathrm{sgn}(d)+\frac{\theta}{2\pi}\right]^2+g(N-3) \left[\frac{\theta}{2 \pi}\right]^2,
		\label{Emeson}
	\end{equation}
	which is notably independent of the distance $|d|$. For $\theta = 0$, the energy simplifies to $\mathcal{E}_\text{meson} = 4m+2g$, as indicated by the line with triangles in Fig.~\ref{Fig3}(a). This implies a transition point
	\begin{equation}
		d_c=2+\frac{2m}{ g \left( 1-\mathrm{sgn}(d)\frac{\theta}{\pi} \right) }.
		\label{dcritical}
	\end{equation} 
	For $|d|<|d_c|$, the string state $|\psi_\text{str}\rangle$ has lower energy, as it involves fewer matter particles compared to the meson state; however, for $|d|>|d_c|$, the meson state becomes more energetically favored. For $\theta=0$, the transition point is where the two curves converge with $|d_c| = 2 + 2m/g$, as shown in Fig.~\ref{Fig3}(a).

	Partial confinement occurs at $\theta = -\pi$, where the string-state energy simplifies to 
	\begin{equation}
		\mathcal{E}_\text{str}^{\theta=-\pi} = 
		\left\{ 
		\begin{aligned}
			&2m+g\left[2d+(N-1)/4\right]  &\text{   for   } d>0 \\
			&2m+g\left[(N-1)/4\right] &\text{   for   } d<0
		\end{aligned}
		\right. .
	\end{equation} 
	Accordingly, the string tension is $\rho_\text{str} = 2g$ for $d>0$, and $\rho_\text{str} = 0$ otherwise. It is clearly indicated that the confining effect only occurs for $d>0$. For $d<0$, $\mathcal{E}_\text{str}$ is independent of $d$, suggesting a deconfinement with $d_c=\infty$. This phenomenon, that the confining property depends on the relative positions of the opposite charges, is termed the partial confinement. Visually, as illustrated in Fig.~\ref{Fig2}(b), the string states for $d>0$ and $d<0$ are subjected to different electric potentials. For the former, the energy density within the string is $g(3/2)^2$, while that of the vacuum is $g(1/2)^2$, resulting in a non-vanishing string tension $\rho_{\text{str}}=  2g$, which is twice the value corresponding to the $\theta=0$ case. For the latter, the electric fields inside and outside the string have opposite signs but with the same energy density $g(1/2)^2$, thus making $\mathcal{E}$ being $|d|$-invariant. In Fig.~\ref{Fig3}(b), the line with squares depicts the variation of the string state energy $\mathcal{E}_\text{str}^{\theta = -\pi}$ as a function of $d$, while the line with left-pointing triangles represents the meson state energy $\mathcal{E}_\text{meson}^{\theta = -\pi}$. It is evident that the string breaking can only occur in the confined regime $d>0$, with $d_c = 2+m/g$. The value of $d_c$ is smaller compared to the $\theta=0$ case, as shown in Fig.~\ref{Fig2}(b), due to the larger string tension.

	Partial confinement also occurs at $\theta=\pi$, but the dependence of quantities (such as $\mathcal{E}$ and $\rho$) on the sign of $d$ is opposite to the case of $\theta=-\pi$ case, as illustrated in Fig.~\ref{Fig2}(c) and Fig.~\ref{Fig3}(b) [see lines with diamonds and squares]. The underlying mechanism can be understood in the following way. For the Hamiltonian (\ref{H}), $\theta\rightarrow-\theta$ is equivalent to $S_j^{+}\rightarrow S_{j+1}^{+}, S_j^{z}\rightarrow S_{j+1}^{z}$ and $ \psi_j\rightarrow \psi_{j+1}$, the latter corresponds to the charge conjugation $\mathcal{C}$. As a result, the physics under $\theta=-\pi$ with a given $d$ is reproduced by $\theta=\pi$ with $-d$. This suggests that reversing $\theta$ alone can switch between confinement and deconfinement scenarios without the need to adjust the spatial ordering of charges. It would facilitate experimental observation of partial confinement since tuning the topological angle (namely tuning the external field) is generally more accessible than manipulating the particle positions in practice.
	
	For other cases with $\theta \in (-\pi,\pi)$ and $\theta \neq |\pi|$, the string state is generally confined according to Eq.~(\ref{Estr}), with $\mathcal{E}_\text{str}$ for various $\theta$ being shown in Fig.~\ref{Fig3}(c). $\mathcal{E}_\text{str}$ is asymmetric about $d=0$, with the corresponding string tension being given by $\rho_\text{str} = g[1-(\theta/\pi)\mathrm{sgn}(d)]$. The asymmetry of $\mathcal{E}(d)$ and $\rho_\text{str}(d)$ for $\theta \neq 0$ originates from the breaking of both $\mathcal{C}$ and $\mathcal{P}$ symmetry due to the topological angle $\theta$, where $\mathcal{P}$ is the parity operator acting as $S_{j}^{+}\rightarrow S_{-j-1}^{+}, S_j^{z}\rightarrow S_{-j-1}^{z},  \psi_j\rightarrow (-1)^j\psi_{-j}$ on the Hamiltonian (\ref{H}). In Fig.~\ref{Fig3}(d), we fix $d = \pm 12$ and display the dependence of $\rho$ on $\theta$. One can clearly observe that the partial confinement begins to occur at $\theta=\pm\pi$, where $\rho_\text{str} = 0$. For a larger $\theta$, i.e., $|\theta| > \pi$, the string state becomes deconfined with a negative string tension. This is also intuitive, as a strong background electric field would polarize the charging pair and yield a large dipole moment.
		
	\begin{figure}[b]
		\begin{center}
			\includegraphics[width=0.48 \textwidth]{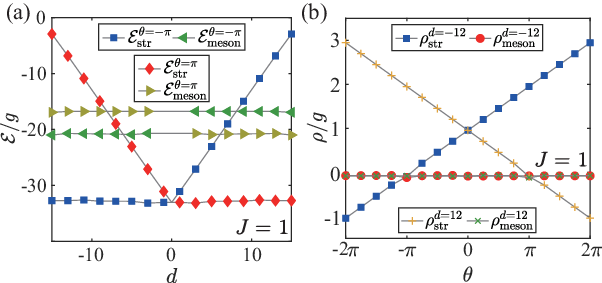}
		\end{center}
		\caption{Equilibrium states' energy $\mathcal{E}$ and string tension $\rho$ in the case of $J=1$, with all other parameters being the same with Fig.~\ref{Fig3}. Panel (a) illustrates the string-like eigenstate energy $\mathcal{E}_\text{str}$ and the meson-like eigenstate energy $\mathcal{E}_\text{meson}$ for $\theta=\pm\pi$. Panel (b) shows the string tension as a function of the topological angle $\theta$ for fixed $d=\pm12$.
		}
		\label{str2}
	\end{figure}
	
	The above results for $J=0$ will not be qualitatively altered when we turn on the matter-gauge interaction $J$. When $J$ is finite, the system lacks integrability, causing a resort to numerical calculations. By setting $J=g=1$ and $m=6g$, we numerically calculate the energy spectrum of the system. Although quantum fluctuations render $s_j^z$ and $n_j$ no longer good quantum numbers, we can still identify low-energy string-like and meson-like states, with the averaged local observables, such as $\langle S_j^z \rangle$ and $\langle \psi_j^\dagger \psi_j \rangle$, resembling the configurations of string and meson states shown in Fig.~\ref{Fig3}(b). Intuitively, the string remains but is thickened by quantum fluctuations. In Fig.~\ref{str2}(a), we present the energy variance of these two eigenstates $\mathcal{E}$ on $d$ for various $\theta$; Additionally, Fig.~\ref{str2}(b) shows the corresponding string tension $\rho$ as a function of $\theta$ for a fixed $d=\pm12$. A comparison between panels Fig.~\ref{Fig3}(b) and Fig.~\ref{str2}(a), as well as Fig.~\ref{Fig3}(d) and Fig.~\ref{str2}(b), clearly demonstrates that the main physical results, such as partial confinement and string breaking, are qualitatively preserved for a non-vanishing $J$.
	
	It may also be necessary to elucidate the differences between the spin-1 QLM discussed here and the spin-1/2 QLM which has been extensively studied both theoretically and experimentally \cite{mil_scalable_2020, zhou2022thermalization, wang_interrelated_2023,cheng2022tunable, BingYangspin-1/2tuningtheta,zhang2023observation, PhysRevLett.109.175302, surace2020lattice, gao2022synthetic, gao2023nonthermal}. The Hamiltonian of the spin-1/2 QLM has the same form as Eq.~(\ref{H}), but with the spin operators $S_j$ being spin-1/2 Pauli operators (up to a constant factor). In this case, even without an external electric field, the string state is no longer well-defined, as there always exist electric strings between the charges (inner string) and outside the charges (outer string). Commonly, the spin-1/2 QLM with the outer string pointing to the right (left) is considered to have an inherent topological angle $\theta=\pi$ ($\theta=-\pi$) \cite{surace2020lattice,cheng2022tunable, BingYangspin-1/2tuningtheta,zhang2023observation}. In contrast to the spin-1 case, the charge conjugation $\mathcal{C}$ now would simultaneously change the order of the charges and the sign of $\theta$, rendering the total energy $\mathcal{E}$ irrelevant to the sign of $d$. In this sense, the spin-1 QLM may serve as a better platform for the study of partial confinement, as it allows for independently changing the charge ordering and $\theta$.

	\section{Partial Confinement in dynamics}  \label{PCDynamics} 
	\subsection{String-state dynamics}
	After discussing equilibrium physics in depth, we now turn to non-equilibrium dynamics. Our objective is to explore whether the physics of partial confinement can be signified by the quantum dynamics out of equilibrium. To this end, we first consider the initial state $|\psi_0\rangle$ to be a string state with the electron and positron residing on the two edges of the chain with $d>0$, i.e., 
	\begin{equation}
		|\psi_0\rangle =|\psi_{\text{str}}\rangle= | 1 _{-1} 0 _1 0  \cdots  0_1 0_ {-1} 1 \rangle.
		\label{psi0}
	\end{equation}
	Notably, this state is an eigenstate of the Hamiltonian $H$ when $J=0$. We then allow the system to evolve under the government of $H$ with $J=g=1$. In our numerical simulations, $N=16$, $m=2g$ are fixed. We primarily focus on the three cases of $\theta=\{-\pi,0,\pi\}$. According to the previous discussions, for such a string configuration $|\psi_{0}\rangle$ with the electron situated to the left of the positron, both $\theta=0$ and $\theta=-\pi$ are confining, while $\theta=\pi$ is deconfining. Since flipping the sign of $\theta$ is equivalent to changing the sign of charge ordering $d$, as discussed in Sec.~\ref{PCEquilibrium}, comparing the dynamics at $\pm\theta$ for a fixed string state can directly signify the partial confinement.
	
		\begin{figure}[t]
		\begin{center}
			\includegraphics[width=.46 \textwidth]{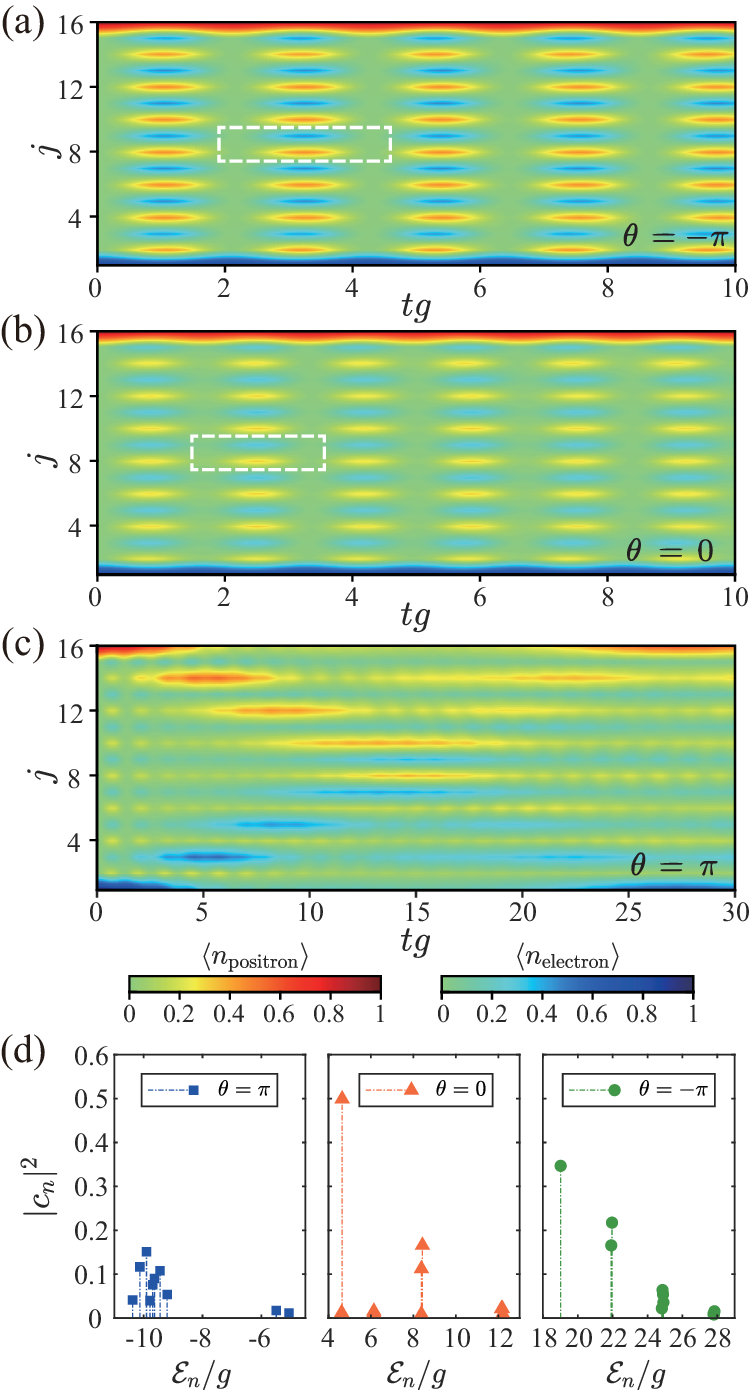}
		\end{center}
		\caption{Density dynamics of the string state. (a)-(c) Dynamics of fermion occupations $\langle n_j\rangle$ for $\theta=-\pi$ [panel (a)], $\theta=0$ [panel (b)], and $\theta=\pi$ [panel (c)], where the occupation status of the positrons and electrons are labeled by the red and blue color bars, respectively, the white boxes indicate the Schwinger mechanism. (d) Top 10 projection probabilities $|c_n|^2$ of the initial state $|\psi_0\rangle$ onto the eigenstates $| n \rangle$ of $H$ under different $\theta$ value. In the calculation, we take $N=16$, $m=2g$, and $J=g=1$.} 
		\label{Fig4}
	\end{figure}
	
	We begin by examining the expectation values of fermion occupations $\langle n_j \rangle = \langle \psi_j^\dagger \psi_j \rangle$ within the time frame $t \leq 30g^{-1}$, as shown in Fig.~\ref{Fig4} with panels (a), (b), and (c) corresponding to the cases of $\theta = -\pi$, $0$, and $\pi$. In the figures, the occupations of positrons and electrons are labeled by red and blue color bars, respectively. A prominent feature is that, for the first two confining cases (i.e., $\theta = -\pi$ and $0$), the edge charges are 'locked' at the boundaries with almost no movement. In the bulk of the chain, $\langle n_j \rangle$ overall exhibits a periodic oscillation, which is indicative of the Schwinger mechanism, as labeled by the white boxes: electron-positron pairs are spontaneously created from the vacuum and then rapidly annihilated with each other. The confinement effect is also evidenced by the small lifetime of the emerged particles (anti-particles) and the fact that they cannot propagate to a wider range on the chain.

	On the other hand, the case with $\theta=\pi$ [Fig.~\ref{Fig4}(c)] exhibits a strikingly different behavior. The two edge charges move towards each other until they meet at the center of the chain at about $t \approx  12g^{-1}$; after that, they reverse their directions and retreat to the boundaries. Unlike the traditional scattering process for free fermions where transmitted waves continue to propagate forward after the scattering event, here we do not observe a clear signal of transmitted wave propagation. Additionally, it can be also notable from the figure that the particle occupation $\langle n_j \rangle$ at the boundaries is significantly decreased compared to the initial state after one round trip. Unlike the Schwinger oscillations in the former two cases, the back-and-forth motion of particles can now only sustain for a few cycles and lacks robust periodicity.
	
	To quantitatively explain the periodicity of $\langle n_j \rangle$ in Fig.~\ref{Fig4}, we calculate the projection probabilities of the initial state, denoted as $|c_n|^2=|\langle n|\psi_0\rangle|^2$, where $|n\rangle$ is the eigenstate of $H$ with energy $\mathcal{E}_n$. In Fig.\ref{Fig4}(d), the largest ten values of $|c_n|^2$ are plotted against $\mathcal{E}_n$, with different $\theta$ values indicated using different markers. For comparison, we have aligned the ground state energy $\mathcal{E}_0$ of different $\theta$ values by shifting the spectra. From the data, it is evident that for $\theta=0$ and $\theta=-\pi$, the initial state $|\psi_0\rangle$ is primarily composed of three high-energy eigenstates. These eigenstates exhibit a definite energy gap $\Delta \mathcal{E}$, which determines the oscillating period of $\langle n_j \rangle$ in the way of $T=2\pi/\Delta \mathcal{E}$. Specifically, for $\theta=0$, $\Delta \mathcal{E}\approx 3.78 g$ and $T\approx 1.66 g^{-1}$, whereas for $\theta=-\pi$, $\Delta \mathcal{E}\approx 2.95 g$ and $T\approx 2.13 g^{-1}$. On the other hand, for the case of $\theta=\pi$, $|\psi_0\rangle$ consists of a broader spectrum of low-energy eigenstates. These states lack a consistent energy gap, which dictates the aperiodic behavior of $\langle n_j \rangle$.
	
		\begin{figure}[pbt]
		\begin{center}
			\includegraphics[width=.46 \textwidth]{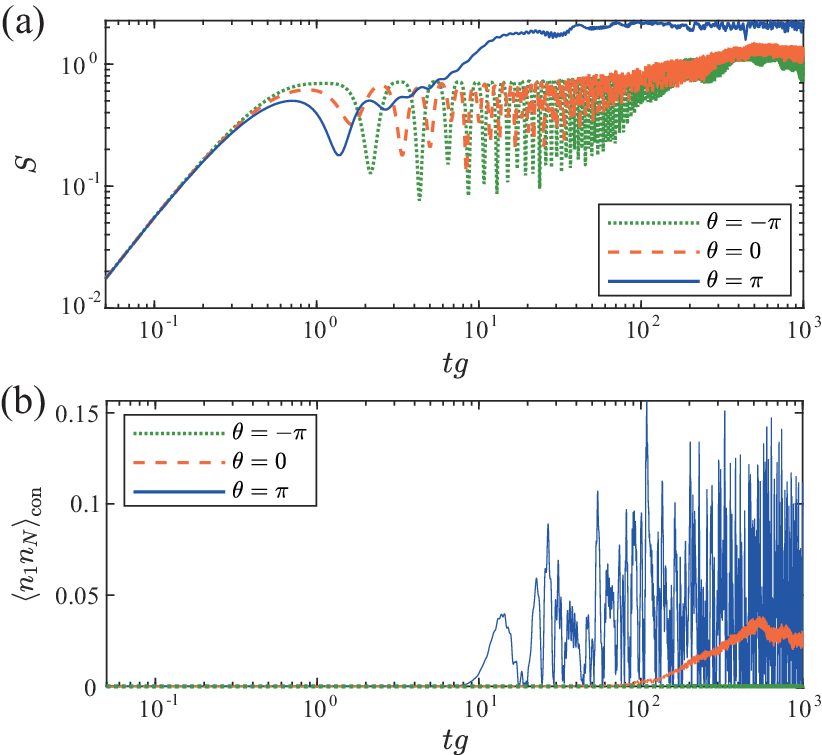}
		\end{center}
		\caption{Dynamics of the entanglement and the correlator for the string state. (a) Dynamics of the bipartite entanglement entropy $S$ for various cases of $\theta$. For the confining cases, $S$ exhibits a periodic oscillations and grows slowly due to Schwinger oscillation. On contrary, for the deconfining case, $S$ presents a rapid increase. (b) Dynamics of the density-density correlator $\langle n_1 n_N \rangle_\text{con}$ between the two boundary fermions, for the confining cases, only weak correlation persisted. Oppositely, for deconfining case, correlation can grow rapidly. The parameters take $N=16$, $m=2J$, and $J=g=1$.}
		\label{Fig5}
	\end{figure}
	
	The confining/deconfining characteristics can also be distinguished from the dynamics of the bipartite entanglement entropy 
	\begin{equation}
		S=-\text{Tr}\rho_\text{R}\ln\rho_\text{R},
	\end{equation}
	where $\rho_\text{R}$ is the reduced density matrix obtained by tracing out the degrees of freedom in one half of the chain. The evolution of $S$ is displayed in Fig.~\ref{Fig5}(a), where the dotted, dashed, and solid lines correspond to $\theta=\{-\pi,0,\pi\}$ respectively. The data reveals that, for the confining cases with $\theta=\{-\pi,0\}$, $S$ exhibits periodic oscillations and grows slowly in the time domain $t>1$; until $t_c\sim 100g^{-1}$, $S$ tends to approach saturation. In contrast, for the deconfining case with $\theta=\pi$, $S$ undergoes a rapid increase in the interval $t \in [1,10]$, following an approximate power-law scaling. It reaches equilibration at $t_c\sim 10g^{-1}$, which is an order of magnitude smaller compared to the former two cases. The time $t_c$ coincides with the moment when the edge charges propagate to the center of the chain [see Fig.~\ref{Fig4}(c)].

	We also calculate the dynamics of the connected density correlation, defined by 
	\begin{equation}
		\langle n_1 n_N \rangle_\text{con} = |\langle \psi_1^\dagger \psi_1 \psi_N^\dagger \psi_N \rangle - \langle \psi_1^\dagger \psi_1\rangle \langle\psi_N^\dagger \psi_N \rangle|,
	\end{equation}
	with the results being presented in Fig.~\ref{Fig5}(b). $\langle n_1 n_N \rangle_\text{con}$ quantifies the density-density correlation between the matter charges on the edges of the chain. Again, various line styles correspond to different cases of $\theta$. It is anticipated that $\langle n_1 n_N \rangle_\text{con}=0$ at $t=0$, since the initial state is a product state. The weak correlation can persist for a considerable duration until $t_c$, beyond which significant correlations between the edge particles begin to develop. For the confining cases with $\theta=\{0,-\pi\}$, $t_c$ is larger than $10^2 g^{-1}$. However, for the deconfining case $\theta=-\pi$, $t_c\sim 10 g^{-1}$, being one order of magnitude smaller than in the previous cases. Therefore, the evolution of edge correlations provides a valuable metric for discerning different confinement statuses.
	
\subsection{Meson-state dynamics}
\begin{figure}[pbt]
	\begin{center}
		\includegraphics[width=.48 \textwidth]{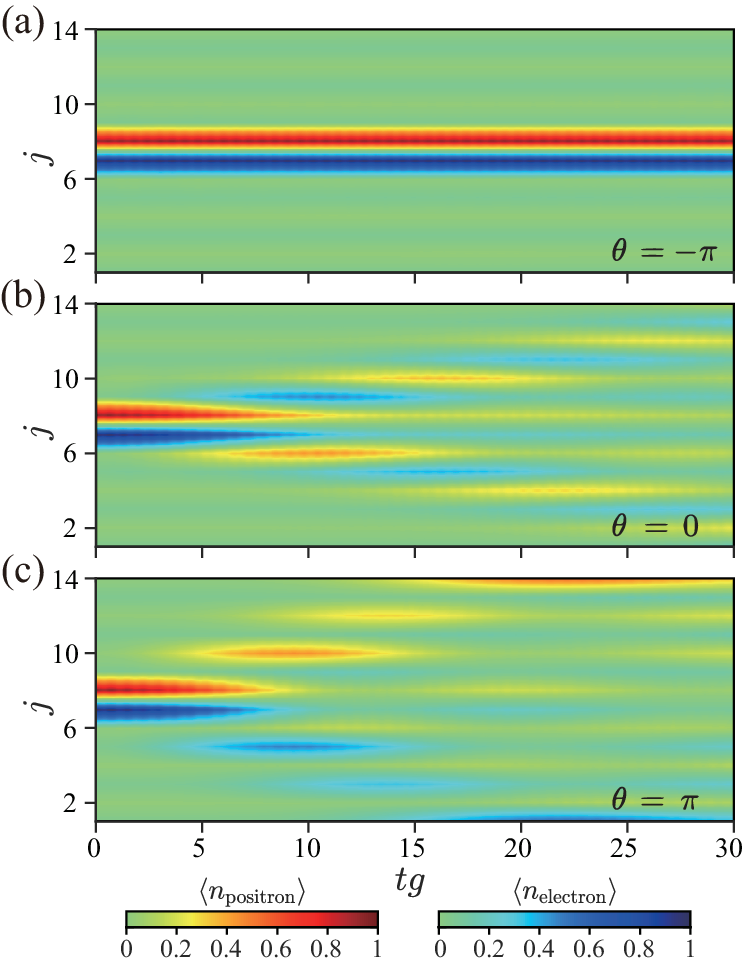}
	\end{center}
	\caption{Density dynamics of the meson state in the cases of $\theta=-\pi$ [panel (a)], $\theta=0$ [panel (b)], and $\theta=\pi$ [panel (c)], where the occupation status of the positrons and electrons are labeled by the red and blue color bars, respectively.
			In the calculation, we take $N=14, m=4g$ and $J=g=1$.}
	\label{scat}
\end{figure}

Up to now, our discussion has focused on the quench dynamics of the string state. Here, we additionally consider a scenario where the initial state exhibits a single meson excitation with $d>0$ at the center of the ground state, i.e.,
	\begin{equation}
		|\psi_0\rangle = |\psi_{\text{meson}}\rangle=\psi^{\dagger}_{\frac{N}{2}} S^{-}_{\frac{N}{2}} \psi^{\dagger}_{\frac{N}{2}+1} |\psi_g \rangle,
		\label{psi02}
	\end{equation}
where the $|\psi_g \rangle$ is the ground state of the Hamiltonian $H$ [Eq.~(\ref{H})] for $m=4g$ and $J=g=1$. The additional gauge-spin flip $S^{-}_{\frac{N}{2}}$ ensures that $|\psi_0\rangle$ remains within the physical gauge sector. Since $d>0$, the particle-antiparticle pair is confined for $\theta = \{-\pi, 0\}$ and deconfined for $\theta = \pi$. We simulate the dynamics of $|\psi_0\rangle$, with the results of the fermion occupations $\langle n_j \rangle$ being presented in Fig.~\ref{scat}. Again, the panels (a)-(c) correspond to the cases of $\theta = \{-\pi, 0, \pi\}$ respectively.

One can observe that $\langle n_j \rangle$ exhibits distinct behaviors for different values of $\theta$. Specifically, in the case of strong confinement at $\theta = -\pi$, the meson state remains stuck in the center of the chain without movement. For the relatively weaker confining case with $\theta = 0$, the meson simultaneously moves towards both ends of the chain, ensuring conservation of momentum. Thus, for these two confining cases, the positron and electron are bound together. In contrast, for the deconfining case of $\theta = \pi$, the meson dissociates into isolated positron and electron, which then independently move away from each other.
	
\begin{figure}[pbt]
		\begin{center}
			\includegraphics[width=.48 \textwidth]{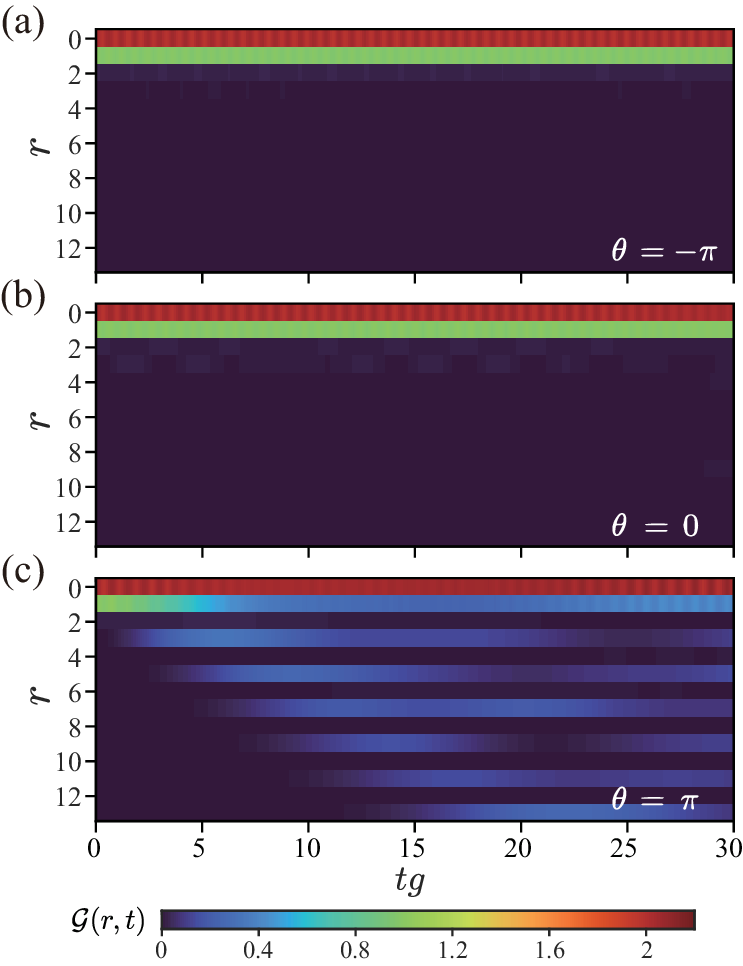}
		\end{center}
		\caption{Dynamics of correlation function for the meson state in the case of $\theta=-\pi$ [panel (a)], $\theta=0$ [panel (b)], and $\theta=\pi$ [panel (c)], with all the parameters being the same with Fig,~\ref{scat}. }
		\label{corre}
	\end{figure}

	The distinction between the confining and deconfining cases can also manifest in the correlation function \cite{cheng2022tunable}
\begin{equation} \mathcal{G}(r,t)=\sum_{j=1}^{N} \langle \psi_0(t)|(n_j-\bar{n}_j)(n_{j+r}-\bar{n}_{j+r})|\psi_0(t)\rangle, \label{cf} \end{equation}
where $n_j=\psi_j^{\dagger} \psi_j$ and $\bar{n}_j=\langle\psi_g|n_j|\psi_g\rangle$. $\mathcal{G}(r,t)$ quantifies the spatial correlation of density fluctuations between any two arbitrary lattice sites separated by a distance $r$. The numerical results of $\mathcal{G}(r,t)$ for various $\theta$ are respectively shown in Fig.~\ref{corre}. 
Specifically, for the confining cases with \(\theta=\{0, -\pi\}\), the correlation function \(\mathcal{G}(r,t)\) remains localized and does not diffuse as $t$ increases. In contrast, for the deconfining case with \(\theta=\pi\), the correlation function between the positron and electron can propagate at a considerable speed and eventually spread throughout the entire space.
		
	\section{Experimental Consideration}  \label{ExperimentalDiss} 
\begin{figure}[pbt]
		\begin{center}
			\includegraphics[width=.5 \textwidth]{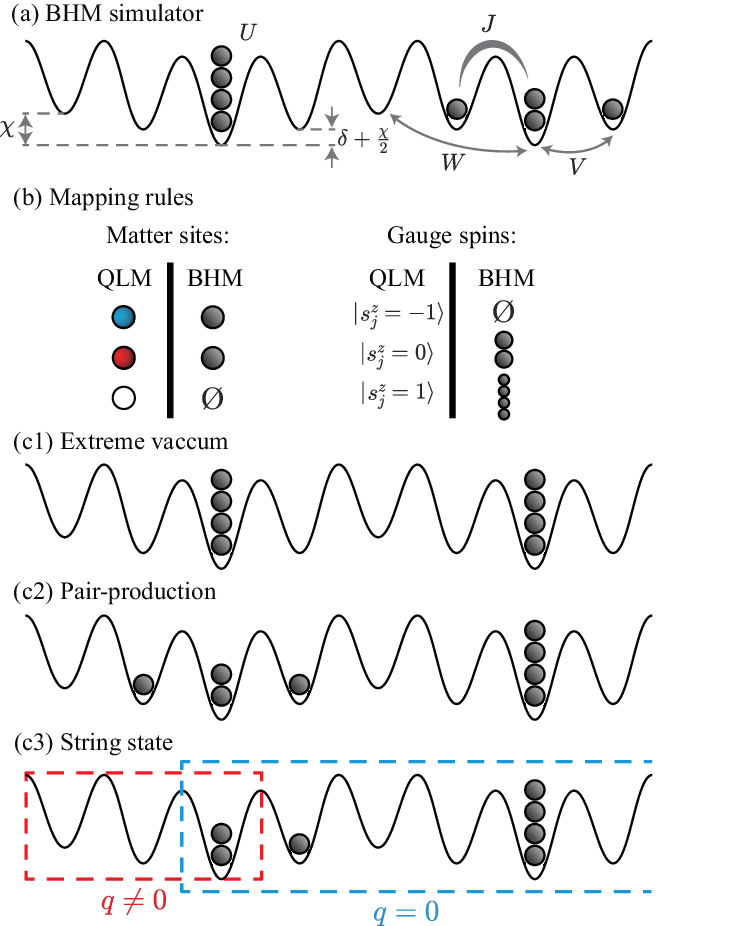}
		\end{center}
		\caption{(a) The schematics of BHM simulator on 1D superlattice, where the tilted potential $\gamma$ is not shown. (b) The mapping relations between the particle representation of the BHM occupation and the particle-antiparticle picture of the QLM. (c) The preparation of the string state, with only the left boundary shown. (c1) The extreme vacuum state. (c2) Local particle-antiparticle pairs are generated using single-site addressing techniques. (c3) The string state is obtained by removing the outer particles. Blue and red boxes respectively indicate the physical and non-physical gauge sectors, with the latter providing a hard-wall boundary for the former.}
		\label{Fig6}
	\end{figure}

	We finally discuss the potential experimental realization using ultracold atoms. The spin-1 QLM has been theoretically proposed to be engineered from a generalized Bose-Hubbard model (BHM) \cite{osborne_spin-s_2023}, which can be implemented with ultracold bosonic gases confined in a superlattice, as illustrated in Fig.~\ref{Fig6}(a). Specifically, the atomic gas is governed by the Hamiltonian
	\begin{align}
		H_{\text{BHM}}&=-\tilde{J}\sum_{l = 1}^{L-1}(b_{l}^{\dagger}b_{l+1}+\text{h.c.}) +\frac{U}{2}\sum^L_{l=1} n_{l}(n_l-1) \notag \\
		&+\sum^L_{l=1}\left[(-1)^{l} \frac{\delta}{2}+l\gamma+\frac{\chi_{l}}{2} \right]n_{l}+V\sum^{L-1}_{l=1}n_{l}n_{l+1} \notag \\
		&+W\sum^{\frac{L}{2}-1}_{l=1} n_{2l-1}n_{2l+1},
		\label{HBHM}
	\end{align}
	where $L=2N$ is the total number of lattice sites, $b_l$ and $b_l^{\dagger}$ are local bosonic operators satisfying $[b_k,b_l^{\dagger}]=\delta_{k,l}$, and $n_l=b_l^{\dagger}b_l$. $\tilde{J}$ is the hopping between neighboring sites, $\delta$ creates energy offsets between matter sites and gauge spins, and $\gamma$ serves as a tilted potential. $\delta$ and $\gamma$ help to eliminate gauge-breaking hoppings. $\chi_{l}$ is a four-site periodic term employed to realize the topological angle \cite{BingYangspin-1/2tuningtheta}, i.e.,
	\begin{equation}
		\chi_l=\left\{\begin{matrix}
			0 \quad & \quad\text { if } \  l \bmod 2=0 \\
			\chi \quad & \quad\text { if } \   l \bmod 4=1 \\
			-\chi \quad & \quad\text { if } \   l \bmod 4=3
		\end{matrix}
		\right. , \label{chi_l}
	\end{equation}
	where $\chi=g {\theta}/{\pi}$. In the second line of Eq.~(\ref{HBHM}),
	$U$ is the on-site interaction, $V$ and $W$ are respectively the nearest-neighbor and next-nearest-neighbor interactions.
	
	The generalized Bose-Hubbard model [Eq.~(\ref{HBHM})] serves as the foundation for realizing the spin-1 QLM [Eq.~(\ref{H})] of length $N$. In this setup, the even lattice sites with $l \in \text{even}$ can only be singly occupied or empty, representing matter particles; whereas the occupancies at odd sites are restricted to $n_{l \in \text{odd}} = \{0,2,4\}$ to realize the three gauge spin states. The detailed mapping relations between the two models are presented in Fig.~\ref{Fig6}(b). To realize the gauge invariance in the $\mathbf{q} = \mathbf{0}$ sector, we focus on the following three configurations: 
	\begin{equation}
		\begin{aligned}
			|C_1\rangle = | 0400\rangle_{\text{BHM}} \ \ \ \leftrightarrow \ \ \ &|0 _1 0 _{-1}\rangle_{\text{QLM}}, \\
			|C_2\rangle = | 1210 \rangle_{\text{BHM}} \ \ \ \leftrightarrow \ \ \ &|1 _0 1 _{-1}\rangle_{\text{QLM}}, \\
			|C_3\rangle = | 0202 \rangle_{\text{BHM}} \ \ \ \leftrightarrow \ \ \ &|0 _0 0 _0\rangle_{\text{QLM}},
		\end{aligned}
	\end{equation}
	where the notation $|n_l, n_{l+1}, n_{l+2}, n_{l+3}\rangle_{\text{BHM}}$ denotes the particle number representation of the BHM occupation status of four consecutive sites for $l \in \text{even}$, and
	the $| n_{j}, ~_{s_{j+1}}, n_{j+2}, ~_{s_{j+3}} \rangle _{\text{QLM}}$ 
	represents the corresponding matter and gauge configuration in the QLM, where $j=l/2$. When $\tilde J = 0$, the bare energy of the three configurations $|C_{1,2,3}\rangle$ are
	\begin{equation}
		\begin{aligned}
			\mathcal{E}_{1} &= 6U-2\delta\pm 2\chi, \\
			\mathcal{E}_{2} &= U+4V\pm\ \chi, \\
			\mathcal{E}_{3} &= 2U+8W-2\delta,
		\end{aligned}
	\end{equation}
	where the $\pm$ determined by the bosonic lattice site index $l$ according to Eq.~(\ref{chi_l}).
	Based on this, all head-to-tail combinations of the three configurations span the entire Hilbert space in the physical sector. 
	
	Notably, there exist configurations that do not preserve gauge invariance. To circumvent these states, it is necessary to ensure that the bare energies of gauge-invariant configurations are nearly resonant, while those of the gauge-violating configurations are far-detuned from resonance. In the case of $U,V,W,\gamma\gg \tilde{J}$, it requires that
	\begin{equation}
		\begin{aligned}
			V\approx & \frac{5U}{4}-\frac{\delta}{2}, \\
			W\approx & -\frac{U}{2}+\delta+2V\approx 2U.
		\end{aligned}
	\end{equation}
	By applying the Schrieffer-Wolff transformation, we can derive the effective Hamiltonian of the Bose-Hubbard model within the gauge-invariant subspace, which takes the form of Eq.~(\ref{H}).
	The detailed coefficient relations are given by
	\begin{subequations}\label{eff_parameters:group}
		\begin{align}
			g&=2U-4W \label{eff_parameters:sub1},\\
			m&=-\frac{3}{2}U+\delta+2V-2W\label{eff_parameters:sub2},\\
			J&=\frac{8\sqrt{12}\tilde{J}^2(2\delta-3U)}{(2\delta-3U)^2-16\gamma^2}, \label{eff_parameters:sub3}
		\end{align}
	\end{subequations}
	where we have omitted the term $\propto \tilde{J}^2$ in Eq.~\eqref{eff_parameters:sub1} and Eq.~\eqref{eff_parameters:sub2} due to the perturbative nature of $J$. Additionally, $\chi$ is considered negligible in the expression of Eq.~\eqref{eff_parameters:sub3} as $|\delta \pm \gamma| \gg |\chi|$ is satisfied \cite{BingYangspin-1/2tuningtheta}.
	
	The extreme vacuum state $|\psi_\text{ev}\rangle$ forms the basis for preparing string states, which is defined by
	\begin{equation}
		|\psi_\text{ev}\rangle =|\cdots0400\cdots\rangle_{\text{BHM}} \leftrightarrow |\cdots0 _{1} 0 _{-1}\cdots\rangle_{\text{QLM}}.
	\end{equation}
	This state, as shown in Fig.~\ref{Fig6}(c1), can be prepared systematically following a well-defined protocol \cite{BingYangspin-1/2tuningtheta,Yang2020NatureObservation,zhang2022functional}. The protocol begins with the system in a uniform superfluid (SF) state. The lattice potential is then gradually modified to establish the desired staggered structure by ramping the parameters $\gamma$ and $\chi$. Following this, the ratio $U/\tilde{J}$ is tuned to induce a phase transition from the SF to a Mott insulator state, where deep lattice sites achieve a four-particle occupancy, shallow sites remain vacant, and even sites have an average occupancy satisfying $0 < \langle n_{l\in \text{even}} \rangle < 4$. Thereafter, spin-selective techniques \cite{zhang2022functional} are applied to selectively remove particles from the even sites, resulting in the formation of the extreme vacuum state.
	
	The string state $|\psi_\text{str}\rangle$ [Eq.~(\ref{psi0})] is distinguished from $|\psi_\text{ev}\rangle$ by modifications only at the boundaries. Employing single-site addressing techniques \cite{BingYangspin-1/2tuningtheta,Yang2020NatureObservation,yang2017spin} enables the generation of particle-antiparticle pairs at the boundaries of the vacuum state, as depicted in Fig.~\ref{Fig6}(c2). Subsequently, by locally removing the outer particles through a laser-induced resonant excitation, the string state can ultimately be obtained, as shown in Fig.~\ref{Fig6}(c3). At the left end of the chain, the left outer gauge sector (enclosed by the red box), with configuration $|012\cdots\rangle_{\text{BHM}} \leftrightarrow |_{-1} 0 _{0}\cdots\rangle_{\text{QLM}}$, does not belong to the physical sector $q=0$. This ensures the rest of the chain (marked by the blue frame) operates consistently within the physical sector and experiences a hard-wall boundary. A similar strategy is also employed at the right end of the chain to achieve the complete particle distribution and boundary condition required for the string state.
	
	\section{Conclusion}  \label{Conclusion} 
	To conclude, we have presented a comprehensive investigation of partial confinement on the platform of the spin-1 quantum link model, a promising platform realizable with cold atoms in optical superlattices. The partial confinement is characterized by a dependency of confinement properties on the spatial arrangement of charged particles, manifested by the asymmetry of the equilibrium energy and the string tension of the string state with respect to the charge ordering. 
	In the non-equilibrium dynamics, both string and meson states exhibit strikingly distinct dynamical features depending on their (de)confining status, as reflected in such quantities as local fermion occupations, bipartite entanglement entropy, and edge charge correlations.
	We have also elucidated that manipulating the topological angle can be an effective proxy for controlling charge ordering, thereby simplifying experimental procedures by obviating the need for direct charge manipulation.
	Given that the quantum link model is amenable to current experimental capabilities, our study offers a strategic avenue for exploring novel physics in gauge theories using state-of-the-art quantum simulators. 
	Furthermore, it may also be interesting to discuss the partial confinement in other forms of lattice gauge theories such as the improved Hamiltonian \cite{gustafson2024,Luo1994}.
	
	\begin{acknowledgments}

		L. C. acknowledges supports from the NSF of China (Grants No. 12174236) and from the fund for the Shanxi 1331 Project. W. Z.  acknowledges supports from the NSF of China (Grants No. GG2030007011 and No. GG2030040453) and Innovation Program for Quantum Science and Technology ( No. 2021ZD0302004).

    \end{acknowledgments}
	
	\bibliography{refs}
	
\end{document}